\documentclass[aps,twocolumn,bibnotes]{revtex4-1}

\usepackage{graphicx}
\usepackage{bm}

\begin{document}

\title{Theoretical model and experimental setup for 3D imaging of cavity vacuum with single atoms confined by a nanohole aperture}
\author{Moonjoo Lee and Kyungwon An}
\affiliation{Department of Physics and Astronomy, Seoul National University, Seoul, 151-747, Korea}

\begin{abstract}
Supplementary information is presented on the recent work by Moonjoo Lee {\em et al.~}on the sub-wavelength 3D imaging of the vacuum-field intensity in a cavity by using single atoms confined by a nanohole aperture as a nanoscopic probe.
\end{abstract}

\date{\today}
\maketitle

The purpose of this article is to provide the supplementary information on the recent work by Moonjoo Lee {\em et al.~}\cite{3d-imaging} on the 3D imaging of cavity vacuum with single atoms confined by a nanohole aperture. A theoretical model describing their experiment is presented and additional information on the experimental setup is provided.
This article is organized as follows. 
We first present the theoretical model in Sec.\ I and show in Sec.\ II that the cavity output is proportional to the vacuum field intensity in the linear regime, where both conditions of the intracavity mean photon number and the Rabi angle much smaller than unity are met.
The procedure for numerical fitting is then discussed in Sec.\ III. 
Finally, we present additional information on the experimental setup in Sec.\ IV.

\section{Theoretical model}

\subsection{Equation of motion} 
A two-level atom interacting with a single mode of a cavity is described by the Jaynes-Cummings Hamiltonian \cite{Jaynes-IEEE63, Filipowicz-PRA86}
\begin{equation}
H=\hbar \omega_{c}a^{\dag}a + \frac{1}{2}\hbar\omega_{a}\sigma_{z}+\mu E_{\text{vac}}(\boldsymbol{r})(\sigma_{+}a+a^{\dag}\sigma_{-}),
\label{eq1}
\end{equation}
where $\omega_{c}$ is the resonance frequency of the cavity field and $\omega_{a}$ is that of the atom. Operators $\sigma_{+}$, $\sigma_{-}$ and $\sigma_{z}$ are the Pauli operators with $\sigma_{+}$($\sigma_{-}$) corresponding to the raising (lowering) operation of atomic states. 

When an atom traverses the cavity mode, the atom-cavity system undergoes a coherent evolution given by $U=\exp(-iH\tau/\hbar)$ during the interaction time $\tau$. Then the cavity-field density operator $\rho_f$ evolves as
\begin{eqnarray}
\rho_f(t+\tau) & = \text{Tr}_a [ U(\tau) \rho_{af}(t) U^{\dag} (\tau) ] \nonumber \\
 & \equiv F(\tau) \rho_f(t) ,
\label{eq2}
\end{eqnarray}
where $\rho_{af}$ represents the atom-field density operator and Tr$_a$ the trace over the atomic states. 
We assume that fully excited single atoms are injected into the cavity repeatedly, and thus only the diagonal elements of $\rho_f$ are involved. 
The above evolution results in a \textit{gain} for the field in that energy is transferred from the initially excited atom to the cavity field. 
The $\textit{loss}$ of the field amplitude due to the cavity decay is treated by
\begin{equation}
\frac{d \rho_f}{d t}= -\frac{1}{2} \kappa (a^{\dag} a \rho_f + \rho_f a^{\dag} a - 2 a \rho_f a^{\dag})\equiv  L \rho_f ,
\label{eq3}
\end{equation} 
where only diagonal elements of $\rho_f$ are coupled to each other.

When the arriving atoms obey Poissonian statistics, which is the case in our experiment, an equation of motion for the average field density operator can be derived in the form of a master equation \cite{Filipowicz-PRA86}
\begin{equation}
\frac{d\rho_f}{d t} = \frac{\langle N \rangle}{\tau}[F(\tau) - I] \rho_f + L \rho_f ,
\label{master-eq}
\end{equation}
where $I$ is the unity operator. The term proportional to $F(\tau)-I$ represents the gain while the second term corresponds to the loss.
The master equation indicates that the steady-state field is obtained when the gain equals the loss. 
The theoretical curves in Fig.\ 2 in the main text were found by numerically solving Eq.\ (\ref{master-eq}) while the gain is averaged over both atomic position spread and velocity distribution. The theoretical curves in Fig.\ 3 were averaged over the atomic velocity distribution only since they are used to fit the data already deconvoluted over the position spread. 
In our analysis, the atomic decay into free space during the interaction time is neglected as in other cavity QED experiments \cite{An-PRL94, Fang-Yen-PRA06, Choi-PRL06,  Seo-PRA10, Meschede-PRL85, Rempe-PRL90, Varcoe-Nature00} since it only amounts to a negligible constant background independent of atomic position in the cavity as discussed below in Sec.\ \ref{sec2}.

\begin{figure}[b]
\includegraphics[width=3.4in]{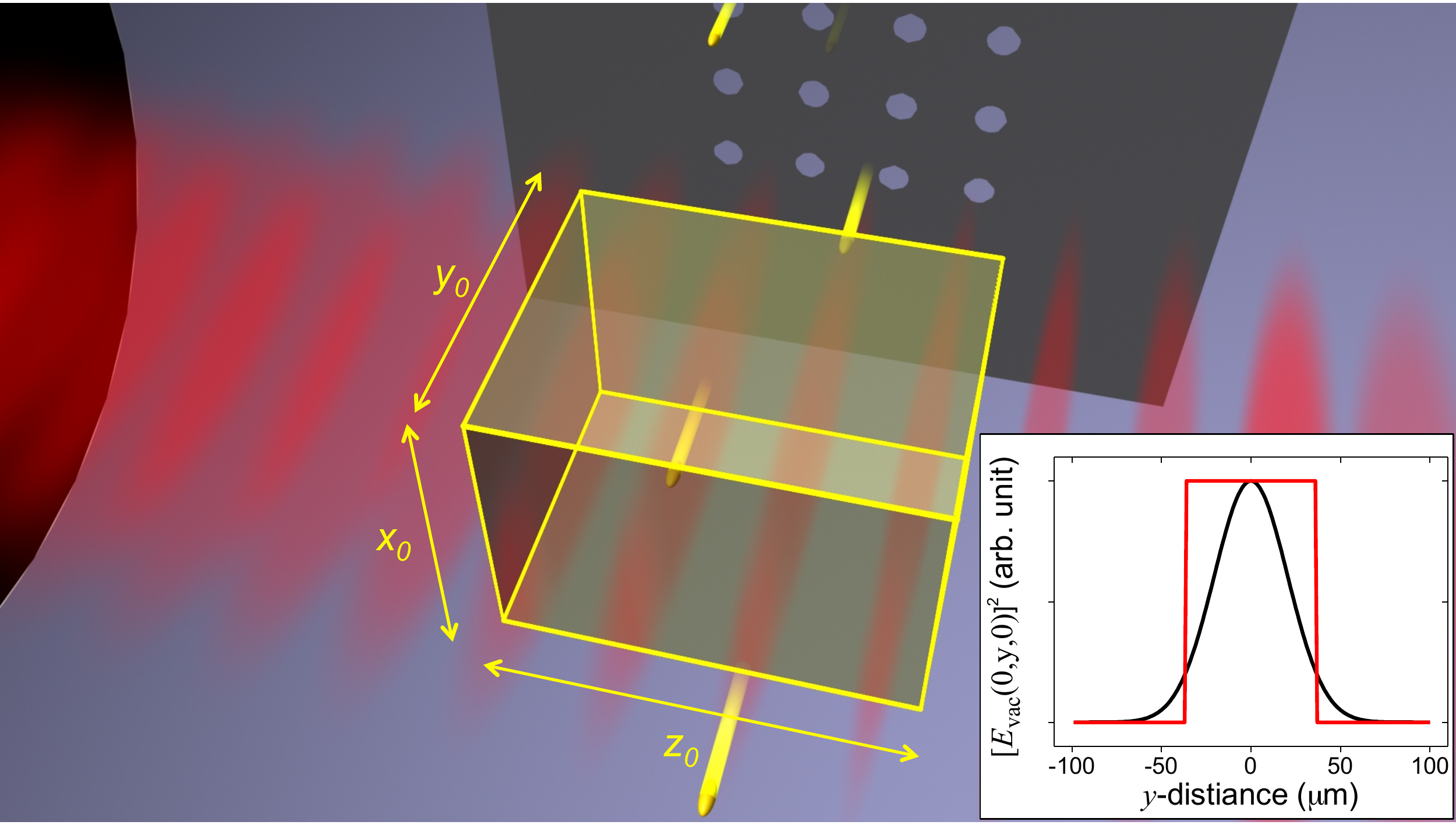}
\caption{A virtual box for defining $\langle N_{\rm tot} \rangle$.
The inset shows a Gaussian mode profile and its associated top-hat profile.
Its $y$ range $y_0=\sqrt{\pi}\omega_{0}$ is chosen such that both top-hat and Gaussian profiles of the cavity field in the $y$ direction result in the same Rabi angle
when an atom traverses these mode profiles.
The actual value of $z_{0}$ is 28.5 $\mu$m and that of $x_0$ is 6.3(12.7) $\mu$m in the linear(nonlinear) regime experiment. 
The illustration is not to scale.}
\end{figure}

\subsection{Definition of $\tau$}
In Eq.\ (\ref{master-eq}), $\langle N \rangle / \tau$ denotes the atomic injection rate. 
In the actual experiment, the cavity mode profile along the $y$ axis, the direction of atomic motion, is a Gaussian (black curve in the inset of Fig.\ S1).
If we take this Gaussian mode profile in the theoretical analysis, we are then faced with an ambiguity in setting the mode boundary or the exact times when an atom enters and exits the cavity mode since a Gaussian extends to infinity.
To resolve this ambiguity we take a top-hat profile (red curve in the inset of Fig.\ S1) instead as was done in most of the previous theoretical treatments \cite{Filipowicz-PRA86, Casagrande-PRL03, Skvarcek-PRA00}. 
The boundary of the top-hat profile with its height the same as that of the Gaussian profile is chosen such that both top-hat and Gaussian profiles in the $y$ direction of the cavity field result in the same Rabi angle
when an atom traverses the cavity mode. The boundary of the top-hat profile is then given by $y=\pm\sqrt{\pi}\omega_{0}/2$. 
The interaction time $\tau$ is then defined as $\tau=\sqrt{\pi}\omega_{0}/v$ with $v$ the velocity of the atom.
 
\subsection{Definition of $\langle N \rangle$}
For defining the mean atom number $\langle N \rangle$, we first define the total atom number $\langle N_{\text{tot}} \rangle$ in the cavity as the averaged total number of atoms in a virtual box whose dimensions are set by the $x$ range $x_0$ and $z$ range $z_0$ of the nanoholes on the aperture and the length $y_0$ of the top-hat profile as shown in Fig.\ S1.  

We then take the imperfect population inversion of the atom into account.
We independently measured the excited state population $\rho_{ee}(t=0)$ of the injected atoms to be $0.86\pm0.04$. 
Here $t=0$ is the time when the atom arrives at $y=-y_0/2=-\sqrt{\pi}\omega_{0}/2$. 
Since the atomic emission probability is proportional to the population inversion, $\rho_{ee}-\rho_{gg}$ with $\rho_{gg}$ the ground state population,
it is reasonable to define the mean atom number as $\langle N \rangle = \langle N_{\text{tot}} \rangle [\rho_{ee}(0) -\rho_{gg}(0) ]$.

\section{Relation between the cavity output and the vacuum field intensity in the linear regime} \label{sec2}

When the atoms are injected into the cavity without atomic polarization, {\em i.e.}, $\rho_{eg}=0$, as in the experiment,
only the diagonal components $\rho_{nn}\equiv p(n)$ of the field density matrix are excited. 
In the steady state, the diagonal elements are obtained as \cite{Filipowicz-PRA86}
\begin{equation}
p(n)=p(0) \prod_{k=1}^{n} \frac{\xi_{k}}{\kappa k},
\end{equation}
where
\begin{equation}
\xi_{k}=\frac{ \langle N \rangle}{\tau} \sin^2(\sqrt{k} \mu E_{\rm{vac}} (x, 0, z) \tau / \hbar ) .
\end{equation}
In the linear regime, we have $\langle n \rangle \ll 1$ and $\mu E_{\rm{vac}} (x, 0, z) \tau / \hbar \ll 1$ (small Rabi angle), and thus
we can approximate $p(n\ge 2)=0$, and consequently we get
\begin{equation}
p(1)=p(0)\frac{\xi_{1}}{\kappa}=\frac{\xi_{1} / \kappa}{1+ \xi_{1} / \kappa}. \\
\end{equation}
Since $\langle n\rangle\simeq p(1)\ll 1$, we should have $\xi_{1} \ll \kappa$.
The cavity output measured in the experiment is just $\kappa\langle n \rangle$, which is then approximated as
\begin{equation}
\kappa \langle n \rangle \simeq \xi_{1} \simeq \frac{\langle N \rangle }{\tau} \left[ \frac{ \mu E_{\rm{vac}}(x, 0, z)}{\hbar}\tau \right]^2,
\label{output-Evac2}
\end{equation}
because of the small Rabi angle condition. 
Equation (\ref{output-Evac2}) shows that in the linear regime the cavity output obtained with the nanohole aperture positioned at $(x,z)$ is proportional to the vacuum field intensity at $(x, 0, z)$. The $xz$ surface plot shown in Fig.\ 2(b) of the main text is based on the equation. 

We can estimate the cavity output contribution by the atomic decay into free space during the interaction time by replacing the Rabi angle factor squared in Eq.\ (\ref{output-Evac2}) with the small population change of about 0.028 during the interaction time and by multiplying the small solid angle of about $10^{-4}$ extended by the cavity mode. The contribution amounts to a negligibly small constant background counts in the experiment.

\section{Quantification of $\boldsymbol{\langle N \rangle}$ and $\boldsymbol{ E_{ \text{vac} } (0) }$ and calibration of $\boldsymbol{\langle n \rangle}$}

\subsection{Basic Idea}

Quantification of $\langle N \rangle$ and $E_{\text{vac}}(\boldsymbol{0})$ and calibration of $\langle n \rangle$ are achieved by numerically fitting the deconvoluted data, such as shown in Fig.\ 3, obtained in the {\em nonlinear} regime with the solutions of the master equation, Eq.\ (\ref{master-eq}). 
It is this nonlinear regime where the previous experiments on one-atom maser/laser \cite{Meschede-PRL85, An-PRL94} were performed. In this regime, $\langle n \rangle$ is no longer much smaller than unity. It can take a value around or greater than unity. 
Because $\langle n \rangle$ is determined by the coherent evolution
where the Rabi angle is proportional to $\sqrt{n+1}E_{\text{vac}}(\boldsymbol{r})\tau$,
as we change the position of the nanoholes and thus $E_{\text{vac}}(\boldsymbol{r})$, the behavior of $\langle n \rangle$ is nonlinear with respect to $\langle N \rangle$ and $E_{\text{vac}} (\boldsymbol{0})$, and thus these parameters can be uniquely determined in this regime.

\subsection{Validity of Fit} \label{validity}

One constraint comes from the validity of the Jaynes-Cummings model. 
When the nanoholes are aligned around the nodes of the vacuum field, the vacuum Rabi frequency 2$g(\boldsymbol{r})$ is smaller than $\kappa$ and $\gamma$.
Although the model can still explain the data qualitatively around those points, it does not provide a quantitatively accurate description and is thus not suitable for determining $\langle N \rangle, E_{\text{vac}}(\boldsymbol{0})$ and $\langle n \rangle$ by fitting.
Therefore, the application of the model for fitting should be restricted to the region of the data where the condition $g(\boldsymbol{r})\gg \kappa, \gamma$ is satisfied. For this, we should use only the data taken while the nanohole aperture is positioned near $x=0$ transversely and scanned along the $z$ axis in a region close to an antinode of the vacuum field.

\subsection{Fitting Model} 

The complete spatial structure is already revealed in the linear regime experiments and we are adapting the top-hat mode along the $y$ axis here. 
Then the interaction Hamiltonian can be written as
\begin{equation}
H_{\rm int}=\mu E_{\text{vac}}(\boldsymbol{0}) e^{-(x/w_{0})^2} \cos{(2\pi z/\lambda)}(\sigma_{+}a+a^{\dag}\sigma_{-})
\label{int-Hamiltonian}
\end{equation}
The interaction is turned on during the interaction time $\tau$ and turned off otherwise. 
Fitting parameters are $\langle N \rangle$, $E_{\text{vac}}(\boldsymbol{0})$, and a scaling factor $\mathcal{S}$ for calibrating the recorded photon counting rate to $\langle n \rangle$. 
Parameters $\langle N \rangle$ and $E_{\text{vac}}(\boldsymbol{0})$ are contained in the master equation, 
and they determine the specific shape of the $\langle n \rangle$ curve as we scan the nanohole aperture.


The fitting method is a linear least squares fit to minimize $\chi^2$ which is given by
\begin{equation}
\chi^2=\sum_{i=1}^{M} \left( y_{i} - \mathcal{S} n_{i} \right) ^2
\end{equation}
where $M$ is the number of the data points to be fitted, $y_{i}$ is the deconvolved photon count, and $n_{i}$ is the numerically calculated mean photon number from the master equation. The values of $\lambda$ and $w_{0}$ in Eq.\ (\ref{int-Hamiltonian}) are already obtained from the sine-squared and Gaussian fits of the data in the linear regime experiments. The other experimental parameters, $\tau$ and $\kappa$, are independently measured. The value of $\mu$ is well known for this atomic transition.

\subsection{Results} 

Figure 3 in the main text shows three data sets to be fitted. They are designated here as $N_1,N_2,N_3$ in order of increasing $\langle N \rangle$.
The data in the inset are the photon counts of the cavity output measured on a single-photon counting module (SPCM) as the nanohole aperture was positioned near $x=0$ and scanned along the $z$ axis from a node to an antinode by $\lambda / 4$ ($\sim$200 nm). The data were then deconvolved with the atomic position spread by using the Richardson-Lucy algorithm and replotted as a function of the relative vacuum field intensity, $[E_{\text{vac}}(0,0,z) / E_{\text{vac}}(\boldsymbol{0})]^2=\cos^2{(kz)}$.
The deconvolved SPCM counts in the main graph would be what we observe if the atoms were injected without any spread in the $xz$ plane. 
By the validity consideration of the Jaynes-Cummings model discussed in Sec.\ \ref{validity}, the deconvolved data in the range of 
$0.5 \le [E_{\text{vac}}(0,0,z) / E_{\text{vac}}(\boldsymbol{0})]^2 \le 1$, where the strong coupling condition is met, were used for the fitting.
The travel distance of the nanohole is about 100 nm along the $z$ axis.

The fit results are shown in Fig.\ 3 and the resulting fit parameters are summarized in Table S1.
We could determine all three fit parameters for the $N_3$ data set. For the $N_2$ date set showing a mild nonlinearity, we used the fit value of $S$ obtained from the  $N_3$ data set since the photo-detection configuration was the same for both $N_2$ and $N_3$ data sets. The fitting then determines the values of $\langle N \rangle$  and $E_{\text{vac}}(\boldsymbol{0})$ uniquely. Note the latter agrees well with that from the $N_3$ data set within the experimental error.

\begin{table} 
\caption{Fitting parameters obtained by the numerical fits.
}
\begin{ruledtabular}
\begin{tabular}{ccccc}
& $\langle N \rangle$ &
\multicolumn{1}{c}{ $E_{\text{vac}}(\boldsymbol{0} )$ (V/cm) }&
\multicolumn{1}{c}{ $\mathcal{S}$ (kcps) }\\
\hline
$N_{2}$&1.1 $\pm$ 0.1 & 0.88 $\pm$ 0.05 &  \\
$N_{3}$&1.5 $\pm$ 0.3 & 0.86 $\pm$ 0.08 & 270 $\pm$ 49\\
\end{tabular}
\end{ruledtabular}
\end{table}

\subsection{Discussions}
\subsubsection{Multi-atom effect} 
The theoretical model in Sec.\ I A or the standard micromaser/microlaser theory describe the interaction between a single atom and a cavity field. 
In our experiment done in the nonlinear regime, the probability of having two or more atoms in the cavity at the same time is not negligible. 
However, it is known from the previous studies that these atoms interact with the common cavity field individually without being affected by the presence of the other atoms if the condition $g(\textbf{r})\tau \ll \pi \sqrt{\langle n\rangle +1}$ is met \cite{ An-JPSJ03}. 
Under this condition, the master equation Eq.\ (S4) can be applied to $\langle N \rangle$ larger than unity. We have confirmed the validity of our numerical calculations based on Eq.\ (S4) with the quantum trajectory simulations \cite{Yang-PRA97, Fang-Yen-OC06} performed for the same $\langle N \rangle$ as shown in Fig.\ 3 in the main text.

\subsubsection{Determination of $\langle N \rangle$ in the linear regime} 
The values of $\langle N \rangle$ in the linear regime experiments were determined as follows;
The parameters $E_{\text{vac}}(\boldsymbol{0})$ and $\mathcal{S}$ are determined from the numerical fits to the data in the nonlinear regime. 
These parameters would remain the same even in the linear regime.
The data in the linear regime are then fitted by the model with only one remaining parameter $\langle N \rangle$, resulting in the quoted $\langle N \rangle$ values.

\section {Additional information on the experimental setup}

\subsection{Nanohole aperture}
The nanohole aperture was made of a 75-nm-thick silicon nitride membrane [250 $\mu$m $\times$ 250 $\mu$m. See Fig.\ 1(d) in the text] with each hole with a diameter of 170 nm milled with the FIB technique (FIB200 by FEI).
Typical milling current and voltage were 1.0 nA and 30.0 kV, respectively. It took about 30 seconds for fabrication of 72$\times$16 nanoholes.
To obtain FIB images, as in Fig.\ 1(d), of nanoholes without damaging them, FIB scanning with 10.0 pA / 30.0 kV was appropriate.

For the experiments done in the linear regime, we used an aperture with 72 $\times$ 16 holes in the $z \times x$ directions, spanning a range of 28.5 $\mu$m $\times$ 6.3 $\mu$m on the membrane. 
For the nonlinear-regime experiment requiring $\langle N \rangle$ to be more than unity, we used a nanohole aperture with the $x$ range of nanoholes extended twice, i.e., 72 $\times$ 32 holes with a hole diameter of 230 nm.
The nanohole aperture assembly was mounted on a nanometre-precision positioner and its $z$ coordinate was tracked in real time by a Michelson interferometer with one arm attached on the aperture mount. 

\begin{figure}
\includegraphics[width=3.4in]{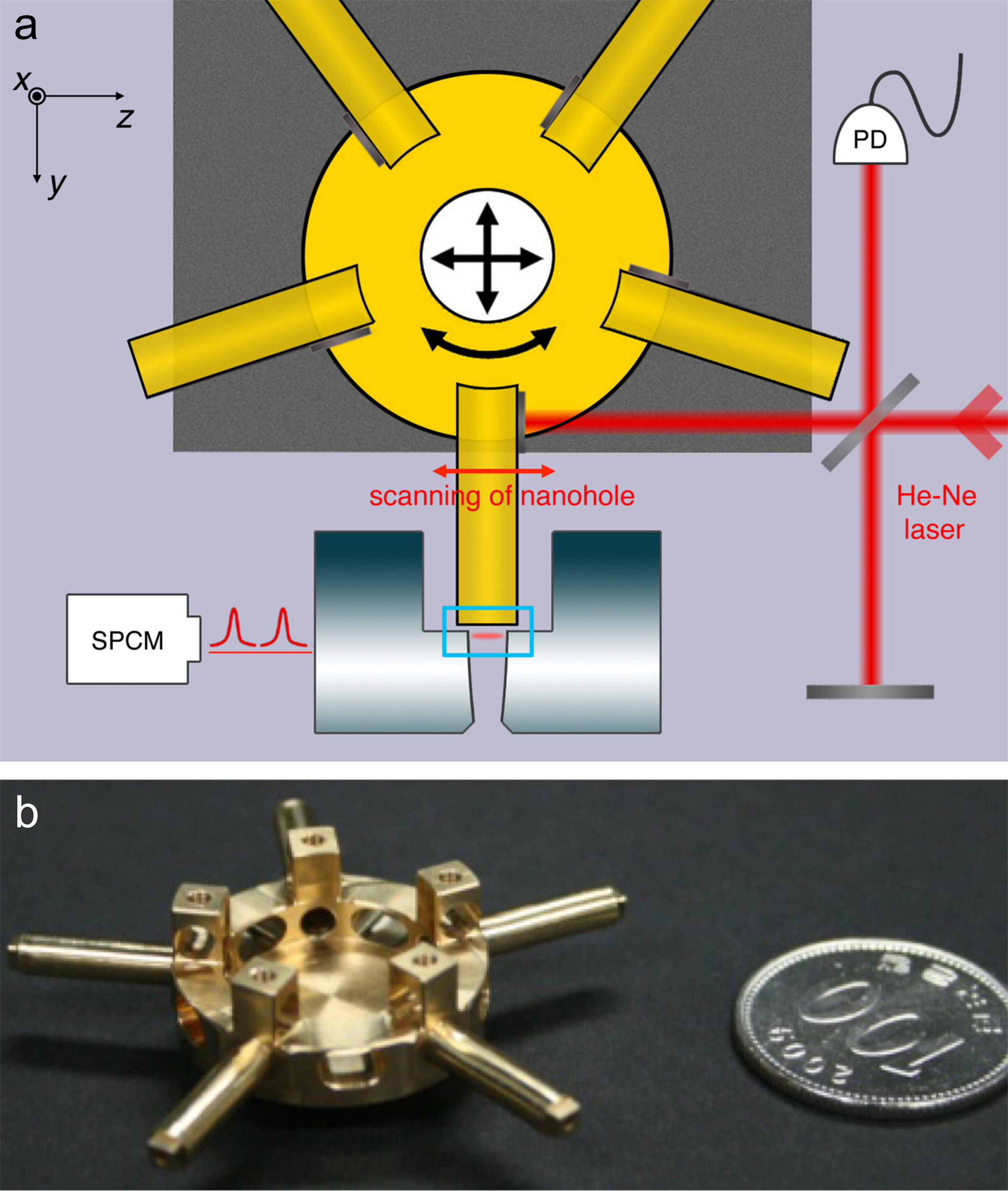} 
\caption{\label{fig:epsart} (a) Experimental setup showing an L-shaped cavity \cite{Kim-OL12}, a revolver for the nanohole apertures and a Michelson interferometer tracking the position of the nanohole aperture in use. PD: photodiode. We present a zoomed-out view of our experiment here whereas only the interaction region (the central blue rectangle) is presented in the main text. (b) Photo of the revolver. The diameter of the central disc is 25 mm, about the size of a Korean 100-Won coin.}
\end{figure}

\subsection{Cavity modification}
The de Broglie wavelength of the atoms injected into the cavity is $3.4\pm0.3$ pm, which guarantees that atomic matter-wave diffraction by the nanoholes is negligible. 
However, the atomic beam itself has a small but finite angular divergence of 0.24 mrad. Unless the nanohole aperture is placed within 500 $\mu$m from the cavity mode axis, atomic position localization with a reasonable contrast cannot be achieved.
We solved this problem by grinding one side of the mirror (7.75 mm in diameter and 8 mm in length) with a diamond wheel in the shape of `L' down to $300$ $\mu$m from the mode without damaging the remaining mirror surface [Fig.\ 1(b)-(c) in the text] \cite{Kim-OL12}.
During the grinding, the mirror surface was covered with a polymer coat for protection.
We found the cavity finesse unchanged from its original value of $1.0\times10^6$.

\subsection{Revolver for the nanohole apertures}
Nanoholes are quickly clogged by atoms under normal operation condition which thus limits our experimental runtime to only 10 minutes. 
To overcome this limitation we have developed a five-arm revolver as shown in Fig.\ S2. 
At the end of each arm, an identical nanohole aperture is mounted. 
When the nanoholes are clogged on the aperture in use, we retreat the revolver, rotate it to the next arm, bring it to the cavity and continue our experiment without opening the vacuum chamber. 
This method extends the overall runtime to about 1 hour.

\subsection{Position tracking interferogram}
Scanning the nanohole aperture in the $z$ direction is done by a shear-type piezoelectric transducer (PZT).
The relative position between the cavity and the aperture is monitored by a Michelson-type interferometer set up as shown in Fig.\ S2(a).
From the interferogram the position of the nanohole aperture can be accurately tracked. 
The mechanical jitter determined by the interferogram is 5$\sim$10 nm, which is small enough for the present experiment.

\subsection{Feed-forward cavity locking}
The cavity resonance frequency slowly drifts away due to continuous heating by thermal radiation from the atomic beam oven and others.
The drift rate at free run is almost constant at about 500 kHz/s during the experiment. 
By applying an additional voltage ramp (1$\sim$1.5 mV/s) to the cavity PZT in the opposite direction to the drift, 
we could keep the cavity frequency on resonance to that of the atom within 100 kHz for more than 100 seconds.
The atom-cavity interaction is not affected by this small frequency mismatch because the transit-time broadening in their interaction is much larger (7.5 MHz).

\newpage

\bibliography{QFLL2}
\end{document}